# Stabilization of Broad Area Semiconductor Laser sources by simultaneous index and pump modulations

W. W. Ahmed[1,2], S. Kumar[1], J. Medina[1], M. Botey[1], R. Herrero[1] and K. Staliunas[1,3]

[1]*Departament de Física, Universitat Politècnica de Catalunya (UPC), Colom 11, E-08222 Terrassa, Barcelona, Spain*
[2]*European Laboratory for Non-linear Spectroscopy (LENS), Sesto Fiorentino 50019, Florence, Italy*
[3]*Institució Catalana de Recerca i Estudis Avançats (ICREA), Passeig Lluís Companys 23, E-08010, Barcelona, Spain*

We show that the emission of broad area semiconductor amplifiers and lasers can be efficiently stabilized by introducing, two-dimensional periodic modulations simultaneously on both the refractive index and the pump (gain-loss) profiles, in the transverse and longitudinal directions. The interplay between such index and gain-loss modulations efficiently suppresses the pattern forming instabilities, leading to highly stable and bright narrow output beams from such sources. We also determine the stabilization performance of the device as a function of pump current and linewidth enhancement factor.

Semiconductor lasers and amplifiers are promising and reliable light sources due to their compact geometry and high conversion efficiency. Despite those advantages, they suffer from a major drawback limiting their applications: the spatially-incoherent light emission at high power regimes. While the use of considerable narrow stripe width (50 $\mu m$ or less) somehow solves the problem of spatial beam quality, it reduces power. High power regimes require broad stripes (200 $\mu m$ or more), and this inevitably leads to low spatial quality of emitted beams. Therefore, in Broad Area Semiconductor (BAS) lasers, the transverse mode instabilities break up the mode profile into multiple filaments, deteriorating the beam quality of high power BAS lasers [1]. Filamentation phenomenon occurs due to the carrier-induced index change effect which relies on the gain saturation and the spatial hole burning [2,3]. Moreover, strong nonlinear interactions of the optical field with the active media lead to complex spatiotemporal dynamics [4].

Several theoretical and experimental attempts to improve the spatial structure of BAS laser have been proposed, such as: structured delayed optical feedback [5], phase conjugated feedback [6], external optical injection [7], a Fourier-optical 4f setup [8] and inward bent curvature in laser facet [9]. In all these techniques, the use of external elements compromises the device robustness while it becomes less compact. It has been found that the origin of BAS lasers' complex spatiotemporal dynamics is in large part caused by intrinsic instabilities [10], which result in pattern formation due to the growth of spatial (transverse) modes. Modulation Instability (MI) is the at the basis of spontaneous spatial pattern formation in many spatially extended nonlinear systems [11]. Pattern formation is a universal phenomenon [12,13], occurring through several classes of instabilities like: the Faraday instability [14], the Benjamin-Feir instability [15], the dissipative parametric instability [16], all extensively studied in optical systems. Particularly, the unstable growing modes generate chaotic pattern dynamics in BAS amplifiers and lasers. As a result, the performance of such high-power lasers sources is severely degraded hindering fiber coupling and restricting their functionality. Thus, compact and efficient schemes to control the detrimental spatiotemporal dynamics are needed to obtain a stable and high-quality output emission from such devices.

In last few years, different schemes based on dispersion management have been reported to improve the spatial structure of BAS amplifiers and lasers. It has been suggested that, in a linear amplification regime, introducing a periodic modulation on the pump profile leads to the spatial filtering of the output radiations [17,18]. However, the situation is more engaged in highly nonlinear regimes due to the appearance of instabilities. More recently, a periodic spatiotemporal modulation of the potential was proposed to manipulate and control MI in a broad class of spatially extended nonlinear dynamical systems [19]. The method has shown to improve the emission of BAS amplifiers and flat mirror VECSELs [20,21]. However, MI is only partially suppressed for pump- modulated BAS amplifiers and also limited to small linewidth enhancement factors (Henry factor), $h$. The periodic modulation of the pump profile induces a combined gain and refractive index modulation, which may prevent full stabilization. It is well-known that two-dimensional (2D) index modulations allow tailoring diffraction, while systems with gain or loss are affected by diffusion. Therefore, it may be expected that simultaneously introducing independent 2D periodic modulations on both the refractive index and the pump of BAS systems, may allow tailoring the system in a more flexible way. Such double-modulation scheme may be able to engineer the diffractive and diffusive properties of the device, eventually leading to stabilization.

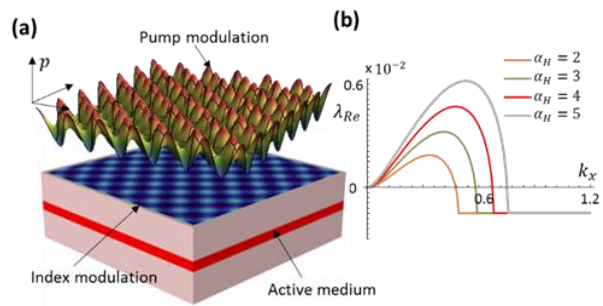

FIG. 1. (a) Schematic illustration of a BAS source with pump (gain/loss) and index modulations. (b, c) Lyapunov spectra of the unmodulated BAS amplifier showing the range of unstable wavenumbers for different linewidth enhancement factors, for a pump $p$=1.2, in Eq. (1).

In this letter, we propose to introduce both, refractive index and gain-loss 2D modulations, in transverse and longitudinal directions, in BAS laser sources which can efficiently and completely stabilize the system. The schematic illustration of the idea with two modulations is shown in Fig. 1(a). In the proposed configuration, the control over the diffraction and diffusive provides a mechanism to obtain bright output beams even for highly nonlinear regimes —note that the unstable spectrum increases with $h$, as shown on Fig. 1(b)—.

We first describe the standard linear stability of an unmodulated BAS laser source. Next, we perform the corresponding Floquet stability analysis to show how the interplay between the 2D periodic modulations of the index and pump profile leads to a complete stabilization of the system. Further, we perform the full numerical integration of the dynamical model to validate the Floquet stability results. In the final part, we evaluate the stabilization performance as a function of the pump current, linewidth enhancement factor, and the pump parameter.

Note that the index and gain/loss modulations can be introduced either in phase or in antiphase. According to recent results [22], phase shifting the modulations of both quadratures could lead to additional stabilization effects relying on non-Hermitian or PT-like potentials, which is beyond the scope of this analysis.

We consider a roundtrip model to describe the field evolution along the BAS laser source [23] as: (see Ref. [27] for a full model analysis):

$$\frac{\partial A}{\partial z} = \frac{i}{2k_0 n}\frac{\partial^2 A}{\partial x^2} + s\left[\frac{p-1}{1+|A|^2}(1-ih) - ih - \alpha\right]A. \quad (1)$$

where $A(x,z)$ is the slowly varying envelop of the complex optical field propagating along the longitudinal direction, $\alpha$ is the linear absorption and scattering loss, $k_0$ is the wavenumber of the incoming light, $n$ is the effective refractive index, $p$ is the pump parameter and $h$ is the Henry factor or linewidth enhancement factor ranges typical from $1 < h < 10$ for semiconductor laser sources [24]. Here $s$ is inversely proportional to light matter interaction length. Typically s is of the order of $\sim 10^{-2} \mu m^{-1}$ and strongly depends on the construction of lasers and semiconductor materials. Note that s can be related with the field confinement factor [25].

To perform the standard linear stability analysis of the unmodulated case, we determine the homogenous state: $A_0 = \sqrt{(p-1)/\alpha - 1}$ and then study the response of the system to a small perturbation around that steady state. We assume that the perturbation of a mode with transverse wavevector, $k_x$, grows (or decays) exponentially. After linearization, we obtain the following expression for the instability spectra:

$$\lambda(k_x) = \frac{1}{2}\text{Re}\left[-c \pm i\sqrt{\frac{k_x^2}{nk_0}\left(\frac{k_x^2}{nk_0} - 2ch\right) - c^2}\right]. \quad (2)$$

where $c = 2s\,\alpha(p-\alpha-1)/(p-1)$. Note that instability spectrum depends on the pump, $p$, and linewidth enhancement factor, $h$. The bandwidth of unstable wavenumbers is $0 < k_x < \sqrt{2hcnk_0}$ which is compatible with Ref. [26].

There are many possibilities to introduce the modulation in such kind of semiconductor devices. Here, we propose to simultaneously introduce a 2D periodic modulation through the pump and the refractive index profiles, in the following way:

$$\frac{\partial A}{\partial z} = \frac{i}{2k_0 n}\frac{\partial^2 A}{\partial x^2} + s\left[\frac{p + 4m_1(x,z) - 1}{1+|A|^2}(1-ih) - ih - \alpha\right]A + ik_0 4m_2(x,z)A. \quad (3)$$

where $m_{1,2}(x,z) = m_{1,2} Cos(q_x x) Cos(q_z z)$, $m_1$ and $m_2$ being the amplitudes of the pump and the refractive index modulations of the same spatial profile and $q_x$ and $q_z$ the corresponding modulation wavenumbers, in the transverse and longitudinal directions. The spatial modulations are assumed in the transverse and longitudinal directions on small spatial scales, i.e. $|q_x| \gg |k_x|$ and $|q_z| \gg |\lambda|$, where $k_x$ and $\lambda$ are the typical transverse wavevector and exponential instability grow parameter. The modulation of the system may be characterized by a geometrical parameter relating both wavenumbers: $Q = 2nk_0 q_z/q_x^2$ where $q_x = 2\pi/d_\perp$, $q_z = 2\pi/d_\parallel$ being $d_\perp$ and $d_\parallel$ denote the transverse and longitudinal period of the modulation, respectively.

According to the Bloch theorem, the steady-state solution of the modulated BAS system can be described by the transverse and longitudinal harmonics $(n, l)$ of the modulations in the form: $A(r) = \sum a_{n,l} e^{i(nq_x x + lq_z z)}$. The interaction between the fundamental mode $a_{0,0}$ and higher order harmonics $a_{1,-1}$, $a_{-1,-1}$, … is in resonance when $Q = 1$, which allows to manipulate the spatial dispersion of the system [19]. Such mutual resonance between the transverse-longitudinal harmonics is expected to suppress the region of unstable wavevectors that excite the chaotic dynamics.

Next, the modified Floquet stability analysis is performed to determine the stability of the periodically doubly-modulated BAS sources. In this method, a set of perturbations is introduced to the steady state solution and the linear evolution matrix of the perturbations is calculated by integrating over one longitudinal period ($d_\parallel = 2\pi/q_z$). The evolution of the diagonal elements of the linear evolution matrix is determined by the real part of the obtained Lyapunov exponents, $\lambda_{Re}(k_x)$. The details of modified Floquet procedure are provided in Ref. [19]. The sign of the Lyapunov exponent indicates the system stability, i.e.; the exponential growth (unstable) or decay (stable) of the perturbation. The parameter space $(m_1, m_2)$ is explored at resonance, $Q = 1$, to plot the stability map as a function of the maximum Lyapunov exponents, $\lambda_{Re,max}(k_x)$, see Fig. 2(a). The white region encircled with a black curve corresponds to the complete stabilization, for which $\lambda_{Re,max}(k_x) \leq 0$. It is evident that after introduction of a small index modulation,

$m_2 \neq 0$, the combined effect of pump and index modulations completely stabilizes the system. The eigenvalues for the representative points, showing complete and partial stabilization of the system, are provided in Fig. 2(b) and Fig. 2(c), respectively. In both cases, the Lyapunov exponents of the modulated system lay below the dotted curve corresponding to the unmodulated unstable case. When all Lyapunov exponents are negative, the system is fully stabilized see Fig. 2(b).

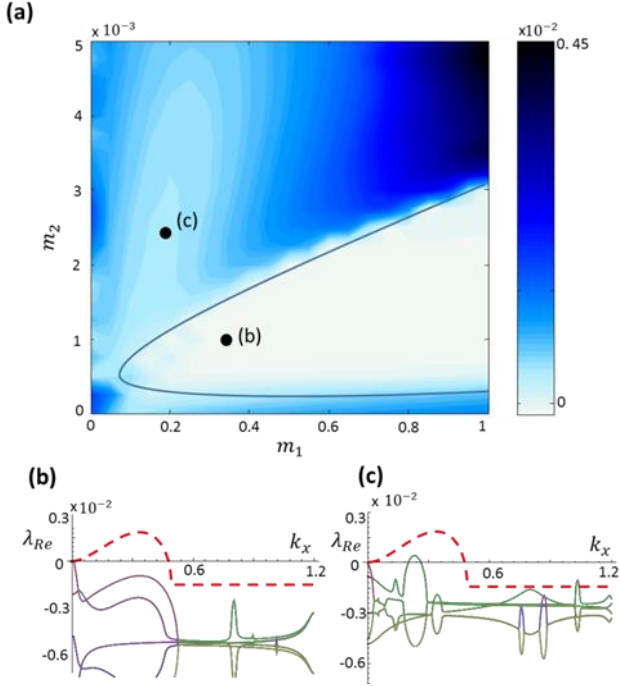

FIG. 2. (a) Stability map obtained from the modified Floquet stability analysis for $h = 2$, $p = 1.2$, $s = 0.03\mu m^{-1}$, $k_0 = 2\pi \mu m^{-1}$, $n = 3.3$ and $\alpha = 0.1$. The real eigenvalues curves, Lyapunov exponents, are shown for complete stabilization in (b) and partial stabilization with remaining instability in (c).

The Floquet stability results are further confirmed by the direct numerical integration of the unmodulated and modulated BAS systems, as shown in Fig. 3(a) and Fig. 3 (b), respectively. The left and right panels, in both cases, represent the field evolution and corresponding angular spectrum during propagation. For the unmodulated case, a chaotic behavior is clearly observed which is also evident from the wide range of unstable wavevectors in the corresponding spectrum. On the other hand, such chaotic behavior may be eliminated by pump and index modulations resulting in a complete stabilization. The spatial dynamics study confirms the results obtained from the Floquet stability analysis, exhibiting a perfect consistency between the predictions and the direct numerical integration of the model. Note that the inspection of the full model (see Ref. [27]) shows a good agreement with the basic results obtained from the simplified model, Eq. (3).

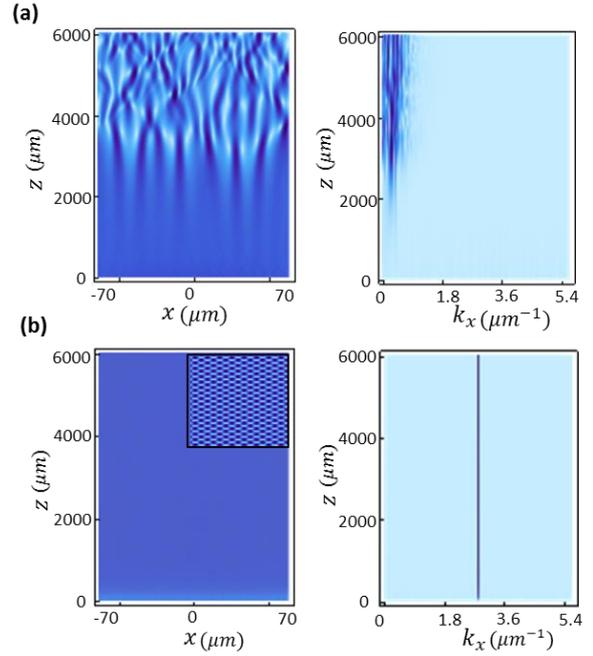

FIG. 3. Direct numerical integration results: (a) unmodulated case, (b) 2D doubly-modulated case ($m_1$=0.35, $m_2$=0.001) exhibiting complete stabilization. In (a,b), left column shows the field profiles and right column shows the corresponding angular spectrum. The zoomed-in view of the field in modulated case is depicted in the inset. The simulation parameters are: width= $140 \mu m$, $h = 2$, $p = 1.2$, $n = 3.3$, $s = 0.03 \mu m^{-1}$, $\alpha = 0.1$, $k_0 = 2\pi \mu m^{-1}$, $d_\perp = 2.2 \, \mu m$ and $d_\parallel = 32 \mu m$.

Finally, we perform a quantitative analysis to determine the stabilization performance of doubly-modulated BAS laser sources. In order to explore the neutral instability curve, we calculate the minimum value of linewidth enhancement factor, $h$, up to which full stabilization is obtained, for different pump levels. We determine the neutral instability curve for different bounds of the pump modulation amplitude, see Fig. 4(a). Note that range of unstable wavenumbers in the spectra increases with the pump above the threshold. The results indicate that $h$ decreases for higher pump currents. The general trend of the neutral instability curve remains same when restricting the pump modulation amplitude. The curve is shifted upwards by increasing the limit of $m_1$ i.e. $m_1 \leq 1$. This restriction has also a physical interpretation, since these systems are always modulated with a weak modulation to avoid additional detrimental spatial and temporal effects. We fix the pump current and study the stabilization performance as a function of $h$. We determine the maximum remaining Lyapunov exponent after stabilization for: only pump modulation (blue curve), simultaneous pump and index modulations (green) and compare them with the unstable case (red dotted curve), see Fig.4(b). We find that, as expected, only introducing a modulation in the pump only allows complete stabilization up to $h \approx 1$, while the stabilization is more robust and efficient for a simultaneous modulation of pump and index profiles, reaching stabilization for substantially high linewidth enhancement factors, up to $h \approx 4.5$. The Lyapunov spectra for high nonlinearities, $h = 4$ and $h = 5$ are provided in Fig. 4(c) and Fig. 4(d), respectively, showing complete and partial stabilization for the doubly-modulated BAS system.

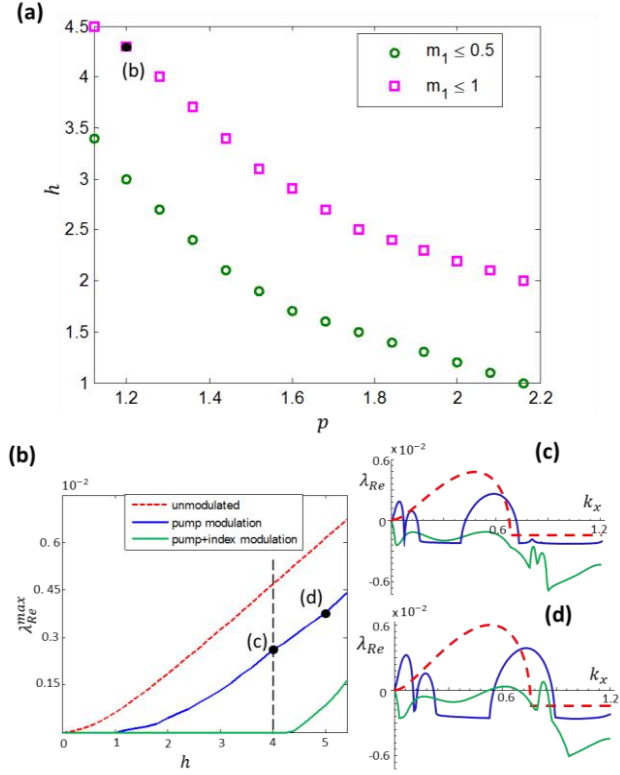

FIG. 4. Stabilization performance of BAS laser sources. (a) Minimum linewidth enhancement factor, $\alpha_H$, for which stabilization is achieved as a function of the pump parameter, by restricting the amplitude of the pump modulation, $m_1$. (b) Comparison of the doubly-modulated, pump plus index, proposed configuration with only pump modulation, as a function of the linewidth enhancement factor for $p=1.2$. (c,d) Lyapunov spectra for $h=4$ and $h=5$ for the unmodulated (red dotted), only pump modulation (blue) and doubly-modulated (green) systems. The stabilization is significantly enhanced in case of pump and index modulation as compared to only pump modulation.

To conclude, we propose that simultaneous 2D modulations of pump and index profiles provide an efficient scheme to stabilize BAS lasers and amplifiers. The proposed configuration can eliminate the MI in highly nonlinear regimes ($h > 4$), with appropriate parameters. We explore the parameter space to determine the MI free regions, showing that best stabilization occurs around the resonance of both modulations, for $Q =1$. Moreover, we perform an analysis of the stabilization performance of the device in terms of the linewidth enhancement factor and the pump parameter for the doubly-modulated BAS laser source. The proposal could be implemented with actual nanofabrication techniques, being moreover compact, contrary to other approaches to improve the emission from BAS laser sources.

## ACKNOWLEDGEMENTS

We acknowledge financial supported by Spanish Ministerio de Economía y Competitividad (project FIS2015-65998-C2-1-P); European Union Horizon 2020 Framework EUROSTARS via project E10524 HIP-Laser and Erasmus Mundus Doctorate Program Europhotonics (Grant No. 159224-1-2009-1-FR-ERA MUNDUS-EMJD).

## References

1. O. Hess, S. W. Koch, and J. V. Moloney, IEEE J. Quantum Electron. **31**, 35 (1995).
2. J. R. Marciante, and G. P. Agrawal, IEEE J. Quantum Electron. **32**, 590 (1996).
3. O. Hess, and T. Kuhn, Prog. Quant. Electr. **20**, 85 (1996).
4. H. Adachihara, O. Hess, E. Abraham, and J. V. Moloney, J. Opt. Soc. Am. B **10**, 496 (1993).
5. S. Wolff and H. Fouckhardt, Opt. Express **7**, 222 (2000).
6. D. H. DeTienne, G. R. Gray, G. P. Agrawal, and D. Lenstra, IEEE J. Quantum Electron. **33**, 838 (1997).
7. A. V. Pakhomov, R. M. Arkhipov, and N. E. Molevich, J. Opt. Soc. Am. B **34**, 756 (2017).
8. S. Wolff, D. Messerschmidt, and H. Fouckhardt, Opt. Express **5**, 32 (1999).
9. J. Salzman, T. Venkatesan, R. Lang, M. Mittelstein, and A. Yariv, Appl. Phys. Lett. **46**, 218 (1985).
10. C. Simmendinger, M. Munkel, and O. Hess, Chaos, Solitons & Fractals **10**, 851 (1999).
11. M. C. Cross and P. C. Hohenberg, Rev. Mod. Phys. **65**, 851 (1993).
12. K. Staliunas, and V. J. Sanchez-Morcillo, "Transverse Patterns in Nonlinear Optical Resonators," Springer Tracts in Modern Physics, Vol. 183, (Springer-Verlag, Berlin, 2003).
13. M. Tlidi, K. Staliunas, K. Panajotov, A. G. Vladimirov and M. G. Clerc, Phil. Trans. R. Soc. A **372**, 20140101 (2014).
14. M. Faraday, Phil. Trans. R. Soc. London **121**, 299 (1831).
15. T. B. Benjamin and J. E. Feir, J. Fluid Mech. **27**, 417 (1967).
16. A. M. Perego, N. Tarasov, D. V. Churkin, S. K. Turitsyn, and K. Staliunas, Phys. Rev. Lett. **116**, 028701 (2016).
17. R. Herrero, M. Botey, M. Radziunas, and K. Staliunas, Opt. Lett. 37, 5253 (2012).
18. M. Radziunas, M. Botey, R. Herrero, and K. Staliunas, Appl. Phys. Lett. **103**, 132101 (2013).
19. S. Kumar, R. Herrero, M. Botey, and K. Staliunas, Sci. Rep. **5**, 13268 (2015).
20. S. Kumar, R. Herrero, M. Botey, and K. Staliunas, Opt. Lett. 39, 5598 (2014).
21. W. W. Ahmed, S. Kumar, R. Herrero, M. Botey, M. Radziunas and K. Staliunas, Phys. Rev. A **92**, 043829 (2015).
22. W. W. Ahmed, R. Herrero, M. Botey, and K. Staliunas, Phys. Rev. A **94**, 053819 (2016).
23. E. A. Ultanir, D. Michaelis, F. Lederer, and G. I. Stegeman, Opt. Lett. 28, 251 (2003)-
24. M. Osinski and J. Buus, IEEE J. Quantum Electron. 23, 9 (1987).
25. G. P. Agrawal, J. Appl. Phys. 56, 3100 (1984).
26. F. Prati and L. Columbo, Phys. Rev. A 75, 053811 (2007).
27. J. Medina, W. W. Ahmed, S. Kumar, R. Herrero, M. Botey, and K. Staliunas, https://arxiv.org/abs/1802.06256, Feb. (2018).